\documentclass[prd,byrevtex
,preprint
,showkeys
,showpacs]{revtex4}
\usepackage{amsmath,amssymb,amsxtra}
\newcommand{\bs}[1]{\boldsymbol{#1}}
\newcommand{\pa}{\partial}
\newcommand{\al}{\alpha}
\newcommand{\del}{\delta}

\begin{document}
\title{The separate universe and the back reaction of long wavelength fluctuations}
\author{Yasusada Nambu}
\affiliation{Department of Physics, Graduate School of Science, Nagoya 
University, Chikusa, Nagoya 464-8602, Japan}
\email{nambu@gravity.phys.nagoya-u.ac.jp}
\preprint{DPNU-05-03}
\date{February 25, 2005 ver 0.9}
\begin{abstract}
  We investigate the backreaction of cosmological long wavelength
  perturbations on the evolution of the Universe. By applying the
  renormalization group method to a Friedmann-Robertson-Walker universe
  with long wavelength fluctuations, we demonstrate that the
  renormalized solution with the backreaction effect is equivalent to
  that of the separate universe. Then, using the effective
  Friedmann equation, we show that
  only non-adiabatic mode of long wavelength fluctuations affects the
  expansion law of the spatially averaged universe.
\end{abstract}
\maketitle
\section{introduction}
The analysis of large scale cosmological perturbation is an important
issue for obtaining information on the initial density fluctuation
that was generated during the era of the inflationary expansion of the
Universe. However, because of the non-linear nature of the Einstein
equation, linear analysis is not sufficient to investigate the
evolution of the early universe. If we consider the the expansion law
of the early universe, the backreaction effect owing to the long
wavelength fluctuations is expected to be important.

Let us consider the Universe with large scale fluctuations of which
wavelength is larger than the Hubble horizon.  Each spatial regions
with the Hubble horizon scale in the Universe are causally
disconnected and evolves independently in time. Hence the Universe
with large scale inhomogeneities can be treated as the collection of
quasi-homogeneous and quasi-isotropic Friedmann-Robertson-Walker (FRW)
universes. The realization of this idea is the separate universe
approach\cite{WandsD:2000} that is equivalent to the lowest order of
the gradient expansion of the Einstein equation. This approach is
suitable to treat a universe with large scale non-linear
inhomogeneities.

An application of this method is the stochastic approach to
inflation\cite{StarobinskiA:1986}.  During the inflationary expansion
of the Universe, long wavelength stochastic fluctuations are generated
and the coarse-grained scalar field in each horizon scale regions
behaves as Brownian particles. The random driving force for the
coarse-grained scalar field appears as the result of the backreaction
of long wavelength quantum fluctuation on the homogeneous
background. Another example of the large scale inhomogeneity that is
tractable by using the separate universe is the preheating stage after
inflation. Long wavelength fluctuations are amplified by the
parametric resonance associated with the oscillation of the background
inflaton field and the super-horizon scale structure of the Universe
evolves to be highly inhomogeneous\cite{TanakaT:2003}.

On the other hand, our present observable universe is considered to be
homogeneous and isotropic, and the evolution of the Universe is
determined by the Friedmann equation. Thus, to incorporate the
backreaction effect into the Friedmann equation, we have to take the
spatial average of an inhomogeneous
universe\cite{MukhanovVM:1997,AbramoLR:1997.1,AbramoLR:1997.2,AbramoLR:1999,
  NambuY:2000,NambuY:2001.1,NambuY:2002.1,GeshnizjaniG:2002,GeshnizjaniG:2003,
BrandenbergerRH:2004}. The
obtained effective Friedmann equation predicts how the expansion law
of the averaged FRW universe is modified by the backreaction
effect. The solution for the cosmological constant problem and the
dark energy problem are investigated in this
direction\cite{AbramoLR:1999,AbramoLR:1997.1,BrandenbergerRH:2004}.

In this paper, we analyze the backreaction effect by long wave
fluctuations using the renormalization group (RG)
approach\cite{NambuY:1999,NambuY:2000,NambuY:2001.1,NambuY:2002.1}. First,
we consider a homogeneous FRW universe with long wavelength linear
fluctuations. We apply the RG method to this system to understand how
the long wavelength fluctuations modify the background FRW
universe. We found that the effect of the backreaction by  long
wavelength fluctuations results in spatially dependent constants of
integration of a FRW universe and the renormalized variables become
solutions of the separate universe approach.  Then, by taking the
spatial average of the solution of the separate universe, we derive
the effective Friedmann equation that involves the backreaction of
long wave modes. The obtained equation shows that the backreaction
effect on the averaged FRW universe appears only for the
non-adiabatic type of fluctuations.


The plan of the paper is as follows. In Sec.~II, we review the
solution of a FRW universe and long wavelength perturbations about
it. Then, the RG method is applied to this system. In Sec.~III,
we derive the effective Friedmann equation by taking the spatial
averaging of the separate universe. Sec.~IV is devoted to summary and conclusion. 
We use units in which $c=\hbar=8\pi G=1$ throughout the paper.

\section{Renormalization of long wavelength mode and the separate universe }
In this section, we apply the renormalization group method to a FRW
universe with long wavelength perturbations and investigate how the
long wave modes modifies the background FRW universe.

\subsection{The solution of  long wavelength fluctuations in a FRW
  universe}

We consider two scalar fields as the matter fields. The metric and the
scalar fields in a flat slice is written as
\begin{align}
  &ds^2=-\left(1-2\frac{\del\!H}{H_0}\right)\frac{d\al^2}{H_0^2}+e^{2\al}d\bs{x}^2, \\
 &\chi^{(1,2)}=\chi_{0}^{(1,2)}(\al)+\del\chi^{(1,2)}(\al,\bs{x}), \notag
\end{align}
where $H_0$ and $\chi_0^{(1,2)}$ are the background quantities, $\del
H$ and $\del\chi^{(1,2)}$ denote the linear perturbation about a
homogeneous FRW universe. In this
slice, the logarithm of the scale factor $\al$ serves as a time
parameter. The background 
Einstein equation and the scalar field equation are
\begin{align}
  &-3H_0^2+\frac{1}{2}\sum_{A=1}^2(\Pi_{0}^{(A)})^2+V(\chi_0^{(A)})=0,
 \notag\\
  &-2H_0H_{0,\al}-3H_0^2-\frac{1}{2}\sum_{A=1}^2(\Pi_{0}^{(A)})^2+V(\chi_{0}^{(A)})=0,
 \label{eq:frw} \\
  &H_0\Pi_{0,\al}^{(A)}+3H_0\Pi_{0}^{(A)}+\frac{\pa
  V}{\pa\chi_{0}^{(A)}}=0,\quad \Pi_{0}^{(A)}\equiv H_0
\chi_{0,\al}^{(A)},\quad A=1,2. \notag
\end{align}
where we have introduced the momentum variable $\Pi_0^{(A)}$ and
${}_{,\al}=\pa/\pa\al$. By using the Hamilton-Jacobi
formalism\cite{SalopekDS:1992}, these equations are combined to the
following equations for the Hubble function $H_0(\chi_0^{(A)})$ and
the scalar fields $\chi_0^{(A)}$:
\begin{align}
  &3H_0^2=2\sum_{A=1}^2\left(\frac{\pa
  H_0}{\pa\chi_{0}^{(A)}}\right)^2+V(\chi_{0}^{(A)}),\label{eq:HJ} \\
  &H_0\chi_{0,\al}^{(A)}=-2\frac{\pa H_0}{\pa\chi_{0}^{(A)}}. \label{eq:HJ-evo}
\end{align}
The solution of Eq.~\eqref{eq:HJ} is written as
$$
 H_0=H_0(\chi_{0}^{(1)},\chi_{0}^{(2)}; d_1, d_2)
$$
where $d_1, d_2$ are integration constants. By differentiating
Eq.~\eqref{eq:HJ} with respect to $d_{1,2}$ and integrating the
resulting equation with respect to the time parameter $\al$, we obtain
the remaining two constants of integration $c_0$ and $f_0$:
$$
 e^{3\al}\frac{\pa H_0}{\pa d_1}\equiv e^{-3c_0}, \qquad 
 e^{3\al}\frac{\pa H_0}{\pa d_2}\equiv e^{-3c_0}f_0.
$$
These four constants of integration  completely specifies the background FRW
universe. Thus the background solution of the FRW universe can be
written as
\begin{equation}
 \chi_{0}^{(1,2)}=\chi_{0}^{(1,2)}(\al+c_0; f_0, d_1,d_2),\qquad
 H_0=H_0(\chi_{0}^{(1)},\chi_0^{(2)},d_1, d_2)
\end{equation}

Equations for long wavelength linear perturbations $\del H$ and $\del\chi^{(A)}$ are
\begin{align}
  & -6H_0\,\del\! H+\sum_{A=1}^2\Pi_{0}^{(A)}\del\!\Pi^{(A)}+\sum_{A=1}^2\frac{\pa
  V}{\pa\chi_{0}^{(A)}}\del\chi^{(A)}=0, \label{eq:hc}  \\
  & \pa_i(\del\!
  H)=-\frac{1}{2}\sum_{A=1}^2\Pi_{0}^{(A)}\pa_i(\del\chi^{(A)}), 
\label{eq:mc} \\
  &-2\left(H_0\del\! H\right)_{,\al}-6H_0\del\!
  H-\sum_{A=1}^2\Pi_{0}^{(A)}\del\!\Pi^{(A)} +\sum_{A=1}^2\frac{\pa
  V}{\pa\chi_{0}^{(A)}}\del\chi^{(A)}=0,   \\
  & H_0\left((\del\!\Pi^{(A)})_{,\al}+3\del\!\Pi^{(A)}\right)
  +\del\! H(\Pi_{0,\al}^{(A)}+3\Pi_{0}^{(A)})+
  \sum_{B=1}^2\frac{\pa^2 V}{\pa\chi_{0}^{(A)}\pa\chi_{0}^{(B)}}\del\chi^{(B)}=0,  \\
 &\del\!\Pi^{(A)}=\frac{\Pi_{0}^{(A)}}{H_0}\del\! H+H_0(\del\chi^{(A)})_{,\al}. 
\end{align}
Eq.~\eqref{eq:hc} is the Hamiltonian constraint and Eq.~\eqref{eq:mc}
is the momentum constraint.  The growing mode solution of the long wavelength
perturbation is obtained by taking  derivative of the background quantities
with respect to the background constants of
integration\cite{NambuY:1998}:
\begin{align}
  & \del\chi^{(A)}=C(\bs{x})\frac{\pa\chi_{0}^{(A)}}{\pa c_0}
  +F(\bs{x})\frac{\pa\chi_{0}^{(A)}}{\pa f_0},\\
  & \del\!
  H=-\frac{1}{2}\sum_{A=1}^2\Pi_{0}^{(A)}\del\chi^{(A)}=\sum_{A=1}^2\frac{\pa
  H_0}{\pa\chi_0^{(A)}}\del\chi^{(A)}, \\
 & \del\!\Pi^{(A)}=(H_0\del\chi^{(A)})_{,\al} ,
\end{align}
where $C(\bs{x})$ and $F(\bs{x})$ are arbitrary functions of spatial
coordinates.  The gauge invariant variable
that corresponds to the spatial curvature perturbation in a comoving
slice is
\begin{equation}
  \mathcal{R}=-C(\bs{x})
-\frac{\sum_A\chi^{(A)}_{0,c_0}\chi^{(A)}_{0,f_0}}{\sum_A(\chi^{(A)}_{0,\al})^2}F(\bs{x}).
\end{equation}
The function $C(\bs{x})$ corresponds to the adiabatic mode of
perturbations and the curvature perturbation owing to this mode is
constant in time. The function $F(\bs{x})$ corresponds to the
non-adiabatic mode of perturbation and this mode results in 
development of the curvature perturbation. 

\subsection{Renormalization of  long wavelength fluctuations}

We apply the RG method to obtain the backreaction of long wavelength
fluctuations on a background FRW universe.  Up to the first order of
perturbations, the solution of the scalar fields is expressed as
\begin{equation}
 \chi^{(A)}(\al,\bs{x})=\chi^{(A)}_0(\al+c_0; f_0)+
[C(\bs{x})-C(\bs{x}_0)]\left(\frac{\pa\chi_{0}^{(A)}}{\pa c_0}\right)
 +[F(\bs{x})-F(\bs{x}_0)]\left(\frac{\pa\chi_{0}^{(A)}}{\pa f_0}\right),
\end{equation}
where we have chosen the functions $C$ and $F$ so that the perturbation
vanishes at a spatial point $\bs{x}=\bs{x}_0$. We regard the
perturbations as the secular terms in the spatial direction and absorb
them into the background constants $c_0$ and $f_0$. For this purpose, we prepare a
renormalization point $\bs{x}_\mu=\bs{x}_0+\mu\, (\bs{x}-\bs{x}_0)$ and
re-define the integration constants as follows :
\begin{equation}
 c_0=c(\mu)+\del c(\mu; 0),\qquad
 f_0=f(\mu)+\del f(\mu;0).
\end{equation}
The counter terms $\del c$ and $\del f$ are chosen to cancel the
$\bs{x}_0$ dependence of the perturbation solution:
\begin{equation}
    \del c+[C(\bs{x}_\mu)-C(\bs{x}_0)]=0,\quad
    \del f+[F(\bs{x}_\mu)-F(\bs{x}_0)]=0. 
\label{eq:counter} 
\end{equation}
This defines the renormalization transformation: the value of the
original constants at the spatial point $\bs{x}_0$ are mapped to the
constants at $\bs{x}_{\mu}$.  
Then the solution of the scalar fields up to the first order becomes
\begin{equation}
  \chi^{(A)}=\chi_0^{(A)}(\al+c(\mu);
  f(\mu))+[C(\bs{x})-C(\bs{x}_\mu)]\chi_{0,\al}^{(A)}
  +[F(\bs{x})-F(\bs{x}_\mu)]\chi_{0,f_0}^{(A)}.
\end{equation}
By assuming that the renormalization transformation defined by
Eq.~\eqref{eq:counter} forms the Lie group,  we can
obtain the RG equation by differentiating Eq.~\eqref{eq:counter} with respect to $\mu$:
\begin{equation}%
 \frac{dc}{d\mu}
=(\bs{x}-\bs{x}_0)\cdot\nabla C,\qquad
 \frac{df}{d\mu}=(\bs{x}-\bs{x}_0)\cdot\nabla F
\end{equation}
and the solution of the RG equation is
\begin{equation}
 c(\mu)=C[\bs{x_0}+\mu(\bs{x}-\bs{x}_0)],\qquad
 f(\mu)=F[\bs{x}_0+\mu(\bs{x}-\bs{x}_0)].
\end{equation}
The renormalized solution is obtained by setting $\mu=1$:
\begin{equation}
  \chi^{(A)}_{\text{ren}}=\left.\chi^{(A)}\right|_{\mu=1}
  =\chi_{0}^{(A)}(\al+C(\bs{x}); F(\bs{x})).
\end{equation}
At the same time, other variables receive the following renormalization
\begin{align}
  &H_0+\del\! H \rightarrow
  H_{\text{ren}}=H_0(\chi^{(A)}_{\text{ren}}), \\
  &\Pi_0^{(A)}+\del\!\Pi^{(A)} \rightarrow
  \Pi_{\text{ren}}^{(A)}=-2\frac{\pa
    H_{\text{ren}}}{\pa\chi^{(A)}_{\text{ren}}}
  =H_{\text{ren}}\chi_{\text{ren},\al}^{(A)},
\end{align}
and the renormalized metric becomes
\begin{equation}
  ds^2=-\frac{d\al^2}{H_{\text{ren}}^2}+e^{2\al}d\bs{x}^2.
\end{equation}
By introducing a new time parameter $t=\int d\al/(NH_\text{ren})$
using an arbitrary lapse function $N(\al,\bs{x})$, 
the metric becomes
\begin{equation}
  ds^2=-N^2(t,\bs{x})dt^2+e^{2\al(t,\bs{x})}d\bs{x}^2,
\end{equation}
and the renormalized variables satisfy the following set of equations:
\begin{align}
  &3H^2=\frac{1}{2}\sum_{A=1}^2\left(\frac{\dot\chi^{(A)}}{N}\right)^2+V(\chi^{(A)}),\quad 
  \frac{\dot\al}{N}=H, \\
  &\pa_iH=-\frac{1}{2}\sum_{A=1}^2\dot\chi^{(A)}\pa_i\chi^{(A)},\\
  &\frac{\dot
    H}{N}=-\frac{1}{2}\sum_{A=1}^2\left(\frac{\dot\chi^{(A)}}{N}\right)^2,
  \\
  & \frac{1}{N}\left(\frac{\dot\chi^{(A)}}{N}\right)\spdot+3H
   \left(\frac{\dot\chi^{(A)}}{N}\right)+\frac{\pa V}{\pa\chi^{(A)}}=0.
\end{align}
These are the basic equations of the separate universe approach (the
lowest order of the gradient expansion).

In the RG approach to the backreaction problem, the effect of the
backreaction by long wavelength fluctuations modifies the background
constants of integration and the constants acquire spatial dependence
associated with long wavelength fluctuations.  The renormalized
solution with the backreaction effect is equivalent to the solution of
the separate universe. Therefore, we can use the separate universe as
 a starting point to derive the spatially averaged Friedmann equation
for an inhomogeneous universe with long wavelength fluctuations.

\section{Effective Friedmann equation}
In this section, we take spatial average of the solution of separate
universe and derive the effective Friedmann equation. The purpose is
to observe how the expansion law of the spatially averaged FRW universe is
modified by the backreaction effect of long wavelength fluctuations.

In a flat slice, the metric and the Hubble function of the separate universe are
\begin{align}
  &ds^2=-\frac{d\al^2}{H^2}+e^{2\al}d\bs{x}^2, \\
  &H=H(\chi^{(A)}(\al+c(\bs{x}), f(\bs{x}))),
\end{align}
where $c(\bs{x})$ and $f(\bs{x})$ are arbitrary functions of the
spatial coordinates.  To proceed the averaging procedure analytically,
we adopt perturbative approach. We expand the solution of the separate
universe about a homogeneous FRW background up to the second order of
perturbation. By replacing $c(\bs{x})\rightarrow c+\del c(\bs{x}),
f(\bs{x})\rightarrow f+\del f(\bs{x})$ and expanding the solution with
respect to $\del c$ and $\del f$, the Hubble function up to the second
order of perturbation becomes
\begin{align}
  H&=H_0+H_1+H_2, \\
  H_0[\al]&=H(\chi^{(A)}(\al+c, f)), \notag\\
  H_1[\al]&=H_{0,c}\del c+H_{0,f}\del f, \notag \\
  H_2[\al]&=\frac{1}{2}H_{0,cc}(\del c)^2+H_{0,cf}\del c\del
  f+\frac{1}{2}H_{0,ff}(\del f)^2.\notag
\end{align}
The metric up to the second order becomes
\begin{equation}
  \label{eq:sepa-metric}
  ds^2=-\left(1-\frac{2H_1}{H_0}-\frac{2H_2}{H_0}
    +3\left(\frac{H_1}{H_0}\right)^2\right)dt^2+e^{2\al(t)}d\bs{x}^2,
\end{equation}
where a time variable $t$ was introduced by
\begin{equation}
  t=\int\frac{d\al}{H_0(\al)}.
\end{equation}
We can obtain the local scale factor by transforming the metric
\eqref{eq:sepa-metric} to a synchronous frame. We define a new time
variable $\tau$ by the following coordinate transformation:
\begin{align}
  &t=\tau+\beta_1(\tau,\bs{x})+\beta_2(\tau,\bs{x}), \\
  &\frac{d\beta_1}{d\tau}=\frac{H_1[\al(\tau)]}{H_0[\al(\tau)]},\quad
 \frac{d\beta_2}{d\tau}=\beta_1\frac{d}{d\tau}\left(\frac{H_1}{H_0}\right)
 +\frac{H_2}{H_0}. \notag
\end{align}
Then we have
\begin{align}
  &ds^2=-d\tau^2+e^{2\al[t(\tau,\bs{x})]}d\bs{x}^2, \label{eq:sepa-syn}
  \\ 
  &\al[t(\tau,\bs{x})]=\al[\tau+\beta_1+\beta_2]\equiv\al_0+\al_1+\al_2, \notag\\
  &\al_0=\al(\tau),\quad
  \al_1=H_0\beta_1,\quad\al_2=H_0\beta_2+\frac{1}{2}
  \left(\frac{d H_0}{d\tau}\right)(\beta_1)^2. \notag
\end{align}
The metric \eqref{eq:sepa-syn} has the same form as that of a flat FRW
universe except its spatial dependence of the scale factor. 
By assuming that the spatial averaging of the first order variables
vanishes $\langle\del c\rangle=\langle\del f\rangle=0$, the spatially
averaged Hubble parameter is
\begin{equation}
  \tilde H\equiv\left\langle\frac{d}{d\tau}\al[t(\tau,\bs{x})]\right\rangle
=H_0[\al(\tau)]+\left\langle\frac{d\al_2}{d\tau}\right\rangle,
\end{equation}
and the Friedmann equation for the spatially averaged scale factor
\begin{equation}
  e^{\tilde\al}\equiv e^{\langle\al\rangle}=e^{\al_0}\left(1+\langle\al_2\rangle\right) 
\end{equation}
is
\begin{equation}
  \label{eq:FRW}
  3\left(\frac{d\tilde\al}{d\tau}\right)^2=3H_0^2+\rho_{\text{BR}}, 
  \quad \rho_{\text{BR}}\equiv
  6H_0\left\langle\frac{d\al_2}{d\tau}\right\rangle.
 \end{equation}
The term $\rho_{\text{BR}}$ represents the modification of the Friedmann
equation due to the long wavelength backreaction effect. Explicit
form of $\rho_{\text{BR}}$ is given by
\begin{align}
  \rho_{\text{BR}} &=3H_0\frac{d}{d\tau}\left[
        H_0\int\frac{d\tau}{H_0}(H_{0,cc}B^2+2H_{,cf}B
          +H_{0,ff})\right]\langle\del f^2\rangle,  \label{eq:rho}\\
    B(\tau)&\equiv H_0\int d\tau\frac{H_{0,f}}{H_0}. \notag
\end{align}
We notice that the expression \eqref{eq:rho} does not contain $\del c$
that is the source of the adiabatic mode of perturbations. Therefore,
for pure adiabatic type of fluctuation $\del c\ne 0, \del f=0$, we
have $\rho_{\text{BR}}=0$. Hence the effective Friedmann equation does
not contain the backreaction terms and long wavelength fluctuations
owing to the adiabatic mode does not alter the expansion of the FRW
universe.  This result is consistent with the previous analysis of the
backreaction problem\cite{NambuY:2000,NambuY:2001.1,NambuY:2002.1};
the backreaction effect appears from $O(k^2)$ in the long wavelength
expansion and there is no backreaction in the long wavelength limit.
For the non-adiabatic type of fluctuation $\del f\neq 0$, we have
non-zero value of $\rho_{\text{BR}}$ and the backreaction of long
wavelength fluctuations modifies the expansion law of the FRW
universe. Although these results were previously obtained for the
Universe with inflationary
expansion\cite{GeshnizjaniG:2002,GeshnizjaniG:2003}, our analysis
shows that we need generally non-adiabatic type of fluctuation to
obtain long wavelength backreaction effect on the spatially averaged
FRW universe. This result is independent of the expansion law of the
background FRW universe.

\section{conclusion}
In this paper, we discussed the backreaction effect owing to
long wavelength fluctuations from two different perspective: One is
renormalization of constants by long wavelength fluctuations. The
backreaction of long wavelength modes leads to the renormalization of
constants contained in the solution of the background FRW
universe. The other is the averaging of the inhomogeneous universe.
From the first perspective, long wavelength mode generates spatial
dependence of constants of a FRW universe and a
homogeneous universe becomes inhomogeneous one owing to the
backreaction effect of long wavelength modes. From the second
perspective, the effective Friedmann equation gets the additional
contribution of the energy density from long wavelength fluctuations
and the expansion law of the averaged universe becomes different from
that of the original background FRW universe.

In this paper, we derived a general formula for $\rho_{\text{BR}}$ but
have not examined what type of the expansion law can be obtained by
the backreaction of long wavelength modes. This subject will be
reported in a separate publication.

\vspace{1cm}
\noindent
{\bf ACKNOWLEDGMENT} \hfill

This work was supported in part by a Grant-In-Aid for Scientific
Research of the Ministry of Education, Science, Sports, and Culture of
Japan (15540265).


\end{document}